\documentclass[11pt]{article}

\usepackage{amsmath}
\usepackage{amsfonts}
\usepackage{amssymb}
\usepackage{graphicx}
\usepackage{hyperref}
\usepackage[margin=2cm]{geometry}
\usepackage{multirow}

\title{Strong coupling between scales in a multi-scalar model of urban dynamics}

\author{Juste Raimbault$^{1,2,3,\ast}$\medskip\\
$^{1}$ Center for Advanced Spatial Analysis, University College London\\
$^{2}$ UPS CNRS 3611 ISC-PIF\\
$^{3}$ UMR CNRS 8504 G{\'e}ographie-cit{\'e}s\medskip\\
$^{\ast}$ \texttt{juste.raimbault@polytechnique.edu}
}

\date{}

\begin{document}

\maketitle

\begin{abstract}
	Urban evolution processes occur at different scales, with intricate interactions between levels and relatively distinct type of processes. To what extent actual urban dynamics include an actual strong coupling between scales, in the sense of both top-down and bottom-up feedbacks, remains an open issue with important practical implications for the sustainable management of territories. We introduce in this paper a multi-scalar simulation model of urban growth, coupling a system of cities interaction model at the macroscopic scale with morphogenesis models for the evolution of urban form at the scale of metropolitan areas. Strong coupling between scales is achieved through an update of model parameters at each scale depending on trajectories at the other scale. The model is applied and explored on synthetic systems of cities. Simulation results show a non-trivial effect of the strong coupling. As a consequence, an optimal action on policy parameters such as containing urban sprawl is shifted. We also run a multi-objective optimization algorithm on the model, showing showing that compromise between scales are captured. Our approach opens new research directions towards more operational urban dynamics models including a strong feedback between scales.
\end{abstract}

\section{Introduction}

The modeling of urban growth and more generally the dynamics of urban systems is central to the design of sustainable territorial policies, through the understanding of past urbanisation processes and the anticipation of future urban trajectories. The design of sustainable future cities requires an historical knowledge of how past cities came to be and evolved \cite{batty2018inventing}. Several models have been proposed at different scales and integrating different dimensions of urban systems, such as models of land-use change at a mesoscopic scale or systems of cities models at a macroscopic scale \cite{pumain2017urban}.

At the scale of a metropolitan area, Land-use Transport Interaction models \cite{wegener2004land} are for example a widely used tool to estimate the dynamics of spatial distributions of activities (mostly residential location and economic activities) in response to an evolution of the accessibility landscape caused by new transportation infrastructures \cite{raimbault:halshs-02265423}. In a similar context, cellular automata models of urban growth or land-use change study more generally land-use transitions with a high spatial resolution, and are mostly data-driven \cite{clarke2007decade}. At the smaller scale of the system of cities, macroscopic models of urban growth have focused on reproducing the distribution of city sizes, either through economic processes as e.g. \cite{gabaix1999zipf}, or from a geographical point of view focusing on interactions between cities \cite{favaro2011gibrat}.

Territorial dynamics, and more particularly urban dynamics, have according to \cite{pumain1997pour} an intrinsic multi-scalar nature, with successive autonomous levels of emergence from individual microscopic agents to the mesoscopic scale of the city and the macroscopic scale of the system of cities. While models at each scale with distinct ontologies are useful to answer their own questions, an explicit account of inter-scale feedbacks, both top-down and bottom-up, would allow testing policies and interventions distributed and differentiated across scales while not neglecting the interactions between scales \cite{wegener2018multi}. Indeed, the need for sustainable territorial policies would imply the construction of multi-scalar models to take simultaneously into account issues associated to each relevant scale \cite{Rozenblat2018,raimbault:halshs-02284933}.

% rq: no def nor clarification of what is meant by "scale"!

Multi-scalar models of urban dynamics are however still at their beginnings. \cite{murcio2015urban} consider population flows at different spatial ranges from the urban area to the country, but does not incorporate distinct ontologies and processes for the different scales. \cite{batty2005agents} however suggests that a similar formalism can be applied to urban processes at different scales. Multi-level statistical models capture some information at imbricated scales \cite{shu2020modelling}, although they can not be used as dynamical simulation models. Similarly, multi-level cellular automata (CA) models for urban growth include factors influencing urban expansion at multiple scales \cite{xu2019directional}. \cite{cheng2003modelling} propose a general framework for such approaches. \cite{torrens2001cellular} suggest that hybrid models coupling CA with other formalisms is a crucial development in the field. \cite{white2006modeling} introduces a CA with variable grid size to account for heterogeneities across scales. \cite{zhu2020dynamic} couple an agent-based model with a CA at multiple scales. \cite{yu2018modeling} embed a local CA into a regional intercity model and a macroscopic potential model. \cite{ford2019multi} couple at different scales an urban development model with a flooding risk model to forecast the future impact of extreme climate events on the London metropolitan region. \cite{xu2020urban} develop an agent-based model of urban expansion with both macro and micro agents. \cite{raimbault:halshs-02013006} suggests that integrating network dynamics at the link level in a macroscopic urban system models is a way to implement a multi-scale model, as done by \cite{raimbault2020hierarchy} which explores hierarchy properties of cities and networks in this context.

In disciplines neighbour to urban modeling, methods have been developed for multi-scale models. For example in spatial epidemiology, \cite{banos2015importance} combines agent-based modeling for local diffusion dynamics with differential equations at the population scale. The NetLogo software for agent-based modeling includes a specific extension for multi-scale modeling \cite{hjorth2020levelspace}. Multi-scale models have also been used for the simulation of crowd dynamics \cite{crociani2016multi}. The study of traffic is also made more accurate by coupling macroscopic and microscopic models \cite{boulet2020coupling}. The management of ecosystems  requires integrating across actors and scales \cite{belem2013organizational}. These non-exhaustive illustrations highlight how the integration of scales is a crucial feature and issue in the understanding of complex systems \cite{chavalarias2009french}.

This paper contributes to the open question of multi-scalar models of urban dynamics by introducing a new simulation model which integrates a strong coupling between the mesoscopic scale and the macroscopic scale. The dynamics within each scale influence the other and reciprocally in an iterative way. More precisely, the model is simple in its components as we focus on the spatial structure of processes rather than on their multi-dimensionality. Therefore, we take into account only population variables, but both at the macroscopic scale of the system of cities and at the mesoscopic scale of the metropolitan area with an urban morphogenesis model. Our contribution is novel regarding previous works in particular regarding the following points: (i) the stylised model explicitly couples distinct scales and ontologies in a strong manner, most models operating only a weak coupling between scales (i.e. no reciprocal and dynamical feedbacks); (ii) the behavior of the model is systematically studied on synthetic systems of cities using model exploration methods.
 
% rq positioning: not large model but simple whih can be understood (cf requiem for large scale models)
 
The rest of this paper is organised as follows: we first describe the model; we then develop its exploration on synthetic systems of cities, and optimization using a genetic algorithm; we finally discuss developments and implications of this work.

% We describe in the following stylized facts justifying the approach => this could (should?) be an important development/basis for the work?

%Strong coupling and multi-scalarity 
%   Distinct processes and objects within each different scale
%   Weak inter-scale coupling, such as progressive resolution for land-use model, does not considers emergence and autonomous scales
% An integrated model, or strongly coupled, would be \textit{a new model extending the two coupled models in the sense that it includes them in some parameter settings or limit conditions}
%Urban multi-scalar complexity must be captured by a strongly coupled model
%no strongly coupled multi-scalar model in the literature (few examples such as  \cite{murcio2015urban} but without distinct ontologies)
%need to however consider ``simple'' models to be able to understand their behavior and extract knowledge from them
%\textbf{Research objective: }
%\textit{Investigate a strong coupling of a simple urban system interaction network at the macroscopic scale with an urban morphogenesis model at the mesoscopic scale}

\section{Multi-scale urban dynamics model}

\subsection{Rationale}

% ! reformulate - this is the CCS abstract

The main characteristic of our model is a strong coupling between the macroscopic scale of a system of cities with the mesoscopic scale of metropolitan areas. We consider simple urban dimensions with population variables only to describe cities, and use stylised processes. The system of cities evolves following a spatial interaction model as described by \cite{raimbault2020indirect}, assuming that interaction flows between cities will increase their attractivity and thus their population growth rate. Population updates are not done directly following migration flows as in stochastic urban models such as in \cite{james2018zipf}, but with the hypothesis of hidden variables (generally scaling with population) capturing for example economic dimensions, which influence population growth rate. At the mesoscopic scale, we consider the spatial distribution of population within an urban area and focus on the growth of urban form. To achieve this, the reaction-diffusion model of urban morphogenesis described by \cite{raimbault2018calibration} allows capturing concentration forces and dispersion forces, and has been shown to reproduce a large range of existing urban forms. Furthermore, \cite{raimbault2020comparison} showed that this model had a good coverage of the morphological space when compared to other models of urban morphogenesis.

\begin{figure}
%  figure: more details?
 \includegraphics[width=\linewidth]{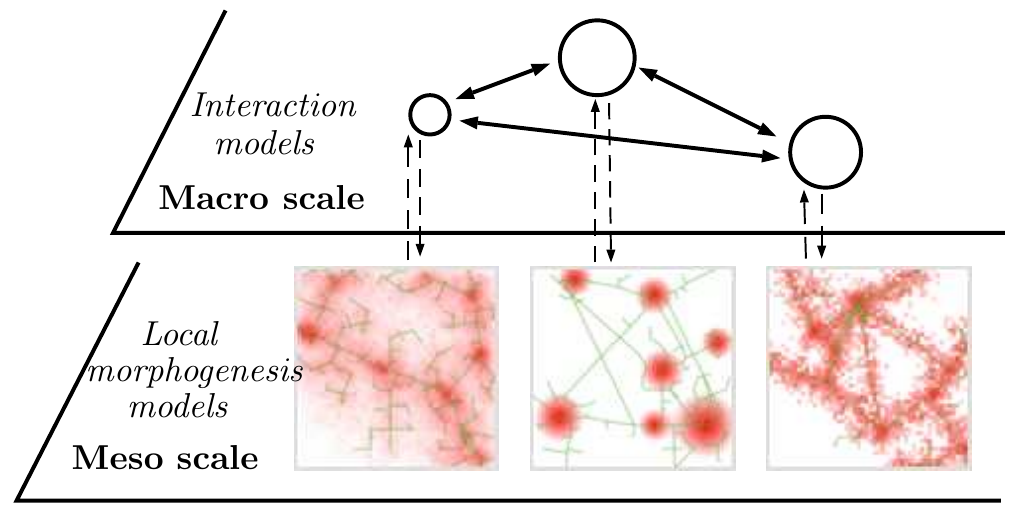}
 \caption{Schematic representation of model architecture. Urban areas with their own urban form are inserted into the macroscopic spatial interaction model, and strong coupling is achieved through top-down and bottom-up feedbacks.\label{fig:fig1}}
\end{figure}

To couple the scales, we assume that (i) macroscopic performance of a city will influence the choices made by planers in terms of land-use, and thus parameters of the morphogenesis mode locally - empirical evidence of such links have been suggested in the literature \cite{joy2015toronto} and we stylise them here; (ii) urban form influences the inner workings of a city through congestion and the fostering of exchanges between agents (individuals, firms), and finally the global insertion of the city - thus spatial interaction parameters are updated as a function of local urban form. This last link is the most discutable in terms of empirical evidence and implies many other dimensions than population only. We however work on a stylised explicative model which must keep a simple structure but on which a systematic knowledge of model behavior can be developed. We show in Fig.~\ref{fig:fig1} a schematic description of model structure.

\subsection{Formalization}

We consider $N$ urban areas, represented at the macroscopic scale by their population $P_j(t)$ at time $t$, and at the mesoscopic scale by a population grid $p_{kl}^{(j)}(t)$. The model runs for a total number $t_f$ of time steps, and we will assume that $\Delta t = 1$ for the sake of simplicity (the formulas can be generalized for arbitrary values of the time step, for example when running on real data with irregular time sampling).

The simulation model operates iteratively, and at each time step goes sequentially through the following steps:
\begin{enumerate}
	\item populations of urban areas are evolved at the macroscopic scale with the spatial interaction model (Equations \ref{eq:eq1} to \ref{eq:eq3} below);
	\item for each urban area, parameters of the moprhogenesis model are updated given parameters capturing typical scenarios: transit-oriented development or sprawl for the diffusion parameter, metropolization or uniformization for the aggregation parameter (Equations \ref{eq:eq4} to \ref{eq:eq6});
	\item local urban form are evolved conditionally to a fixed population increase and following the mescoscopic parameters;
	\item congestions and interactions within areas are synthesised following a cost function, which is used to update the macroscopic interaction parameters (Equations \ref{eq:eq7} and \ref{eq:eq8}).
\end{enumerate}

Details of each stage are described below

\subsubsection{Macroscopic model for populations}

Following \cite{raimbault2020indirect} and \cite{favaro2011gibrat}, aggregated populations variables $P_i(t)$ are evolved according to
	
\begin{equation}
	\label{eq:eq1}
		P_i(t+1) = P_i(t) \left(1 + \Delta t \cdot \left(g_i + \frac{w_i}{N} \cdot \sum_j \frac{V_{ij}}{<V_{ij}>} \right) \right)
\end{equation}
	
The gravity interaction potential $V_{ij}$ between cities is given by 
	
\begin{equation}
	\label{eq:eq2}
		V_{ij} = \left(\frac{P_i P_j}{(\sum_k p_k)^2}\right)^{\gamma_G} \cdot \exp \left(- \frac{d_{ij}}{d_i} \right)
\end{equation}

We write the population variations as
	
\begin{equation}
	\label{eq:eq3}
	\Delta P_i (t) = P_i (t + 1) - P_i (t)
\end{equation}

\subsubsection{Top-down feedback}

The population variations at the macroscopic scale then influence the mesoscopic parameters to capture the top-down feedback. Therefore, the parameters of \cite{raimbault2018calibration}, namely $\beta_i$ of diffusion and $\alpha_i$ of aggregation are both updated. First, the mesoscopic growth rate is adjusted to the population growth uniformly over the time interval such that $N_G^{(i)} (t + 1) = \Delta P_i / t_m$, where $t_m$ is a global model parameter controlling the temporal granularity of local population increases.			
			
The sprawl parameter evolves according to a fixed multiplier and the relative population increase following 

			\begin{equation}
			\label{eq:eq4}
				\beta_i (t+1) = \beta_i (t) \cdot \left(1 + \delta\beta \cdot \frac{\Delta P_i (t)}{\max_k  \Delta P_k (t)}\right)
			\end{equation}
			
where the multiplier parameter $\delta\beta$ allows testing different scenarios: a negative value corresponds to transit-oriented development while a positive value corresponds to an uncontrolled sprawl.
			
The aggregation parameter evolves in a similar way but as a function of accessibility increase, following the rationale that more accessible place with concentrate more activities, as
 
\begin{equation}
	\label{eq:eq5}
	\alpha_i (t+1) = \alpha_i (t) \cdot \left(1 + \delta\alpha \cdot \frac{\Delta Z_i (t)}{\max_k  \Delta Z_k (t)}\right)
\end{equation}
			
where the multiplier parameter $\delta\alpha$ allows switching between a metropolisation scenario (more aggregation) and an uniformisation scenario (less aggregation). In that context, the accessibility is computed as
			
\begin{equation}
	\label{eq:eq5}
	Z_i = \sum_j \frac{P_j}{\sum_k P_k} \cdot \exp( - d_{ij} / d_i)
\end{equation}
			
Change in the level of sprawl ($\beta_i$) depends on the population pressure only, while aggregation depends on accessibility since it is linked to metropolisation processes.

%\textit{Note: the linear scale for these two parameters may not be relevant depending on the distribution of increments ?} $\rightarrow$ to be tested

\subsubsection{Evolution of urban form at the mesoscopic scale}
	
Mesoscopic grids population grids are evolved separately within each urban area, using the updated morphogenesis model parameters. We follow here the same exact steps described in \cite{raimbault2018calibration}, with parameters $N_G^{(i)}$ total population increase, $\beta_i$ diffusion level, $\alpha_i$ aggregation level, $t_m$ time steps and $n_d$ number of diffusions. The model adds iteratively population to grid cells, by distributing it with a probability following a preferential attachment to population, and diffusing it.

In terms of implementation, small population differences in the end between macroscopic and mesoscopic levels (due to rounding in computing the number of steps) is corrected by adjusting the macroscopic increments by the effective mesoscopic increments (which are assumed to be more precise).

\begin{table}
	\caption{Summary of model parameters\label{tab:parameters}}
	\centering
	\begin{tabular}{|c|c|c|c|}
	\hline
		Type & Parameter & Process & Range \\\hline
		\multirow{4}{*}{Macro} & $g_i = g_0$ & Endogenous growth & $\left[0 ; 0.5\right]$ \\
		& $w_i = w_G$ & Interactions weight & $\left[0 ; 0.1\right]$ \\
		& $\gamma_i = \gamma_G$ & Interactions hierarchy & $\left[0 ; 5.0\right]$ \\
		& $d_i$ & Interactions decay & $\left[0 ; 1000\right]$ \\ \hline
		\multirow{4}{*}{Meso} & $\alpha_i$ & Aggregation & $\left[0 ; 5\right]$ \\
		& $\beta_i$ & Diffusion & $\left[0 ; 0.5\right]$ \\
		& $t_m$ & Urban growth speed & $\left[0 ; 20\right]$ \\
		& $n_d$ & Diffusion & $\left[0 ; 5\right]$ \\ \hline
		\multirow{4}{*}{Multiscale} & $\delta\alpha$ & Downward feedback & $\left[-0.5 ; 0.5\right]$ \\
		& $\delta\beta$ & Downward feedback & $\left[-0.5 ; 0.5\right]$ \\
		& $\delta d$ & Upward feedback & $\left[-0.5 ; 0.5\right]$ \\\hline
		& $\lambda$ & Cost of congestion & $\left[0 ; 2\right]$ \\\hline
	\end{tabular}
\end{table}

\subsubsection{Bottom-up feedback}
	
Finally, bottom-up feedback is captured by updating the macroscopic parameters. For the sake of simplicity, only interaction decays are updated, assuming that patterns of urban form play a role in the global insertion of the city. More precisely, we compute gravity flows within the area, and aggregate their value as an economic activity with a squared negative externality interpreted as a congestion with a cost parameter $\lambda$, following
	
\begin{equation}
\label{eq:eq6}
		U_i = \sum_{kl} \left( \frac{P_k P_l}{P^2} \cdot \frac{1}{d_{kl}} - \lambda \left(\frac{P_k P_l}{P^2} \cdot \frac{1}{d_{kl}}\right)^2 \right)
\end{equation}
	
This utility $U_i$ is used to update the interaction decays following
	
\begin{equation}
\label{eq:eq7}
	d_i (t+1) = d_i (t) \left( 1 + \delta d \cdot \frac{U_i}{\max_k \left|U_k\right|} \right)
\end{equation}

where the multiplier parameter $\delta d$ allows controlling for the influence of local performance on global insertion.

\subsection{Parameters}

The Table~\ref{tab:parameters} summarizes model parameters. For clarity, we do not include parameters linked to synthetic model setup, and distinguish them by the scale to which they are attached.

% ! provide nominal values

\subsection{Synthetic setup}

The system is initialized with synthetic systems of cities. A number $N=20$ cities are distributed randomly into a regional urban system, as a square world of width $w=500km$ (reference unit for the decay parameter). City populations follow a Zipf law with hierarchy parameter $\alpha_0 = 1$ (may be modified in experiments), and $P_0 (0) = 100,000$ the initial size of the largest city. Within each city, we consider initial population grids as monocentric (grid of size $W=50$ and center cell density $P_m = 1000$ population units, following an exponential kernel of width one fifth of the world). %check param + necessary experiment: influecne of other initial configs: at least kernel mixture?
The model is simulated for 20 macroscopic time steps (order of magnitude of half a century), each representing $t_m$ mesoscopic time steps (parameter changed in experiments).

%%%
% ! fig placement

\begin{figure}[h!]
	\includegraphics[width=\linewidth]{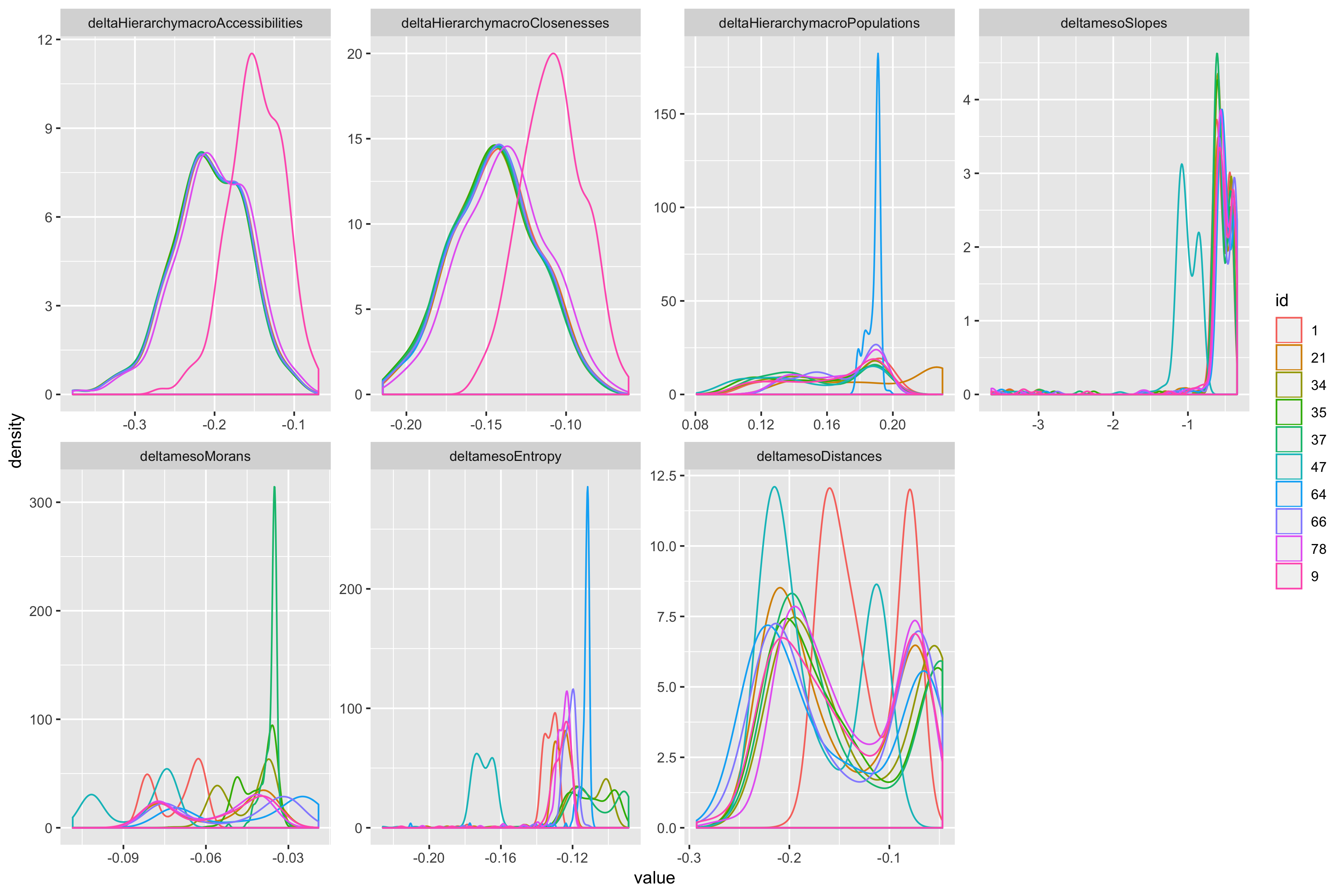}
	\caption{Statistical distribution of indicators for a sample of 10 parameter values.\label{fig:fig2}}
\end{figure}

\begin{figure}[h!]
	\includegraphics[width=\linewidth]{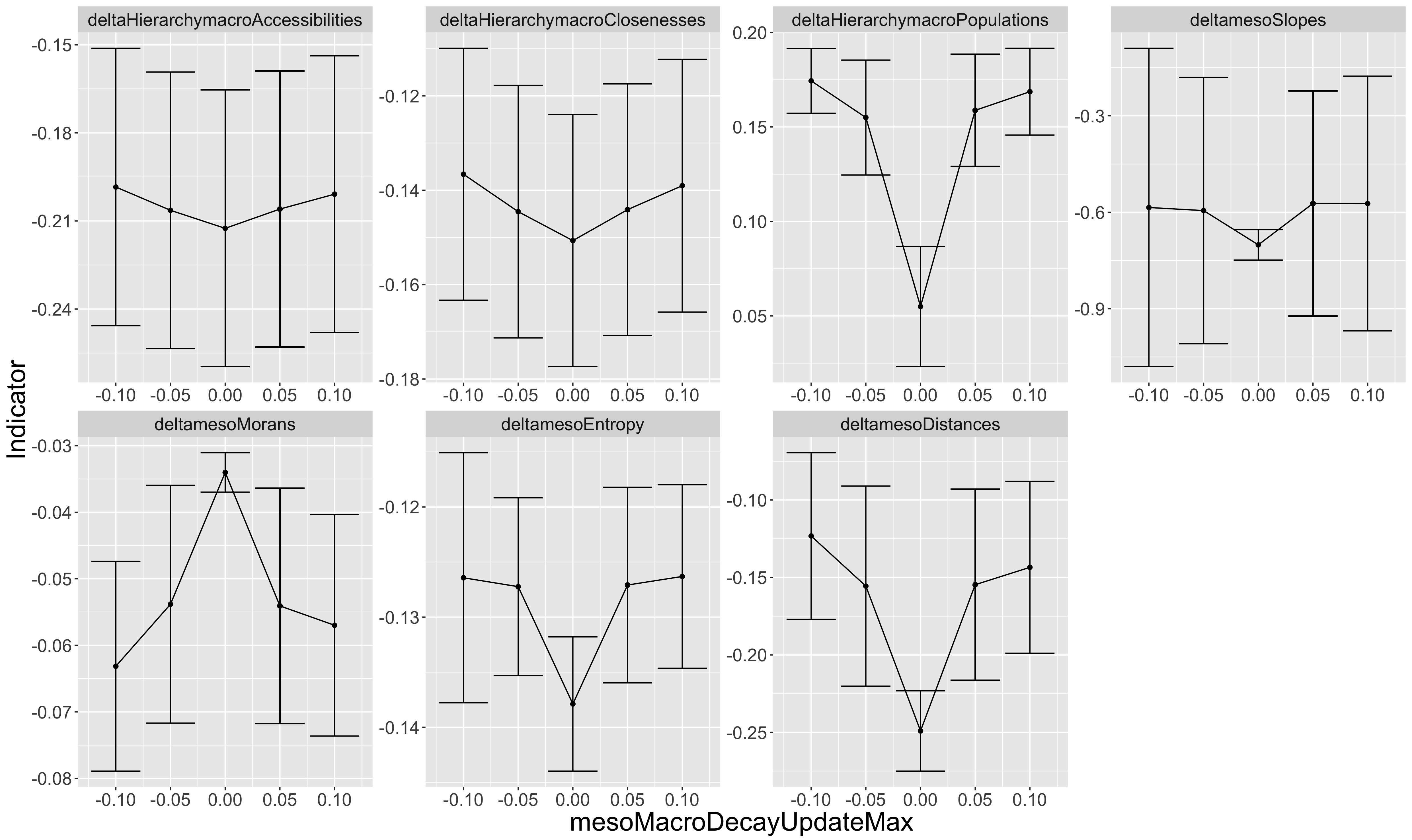}
	\caption{Variations of indicators as a function of $\delta d$, when all other parameters are fixed to their nominal value.\label{fig:fig3}}
\end{figure}

\subsection{Indicators}

Model behavior is characterized using the following indicators: at the  macroscopic scale: distributions of population, accessibilities, and centralities (summarized by average, hierarchy, entropy), following \cite{raimbault2020unveiling}. At the mesoscopic scale, urban form is captured following the indicators used by \cite{raimbault2018calibration}, i.e. Moran index, average distance between individuals, hierarchy of population distribution, and entropy of population distribution.

\section{Results}

\subsection{Implementation}

Model implementation is done under some performance constraints: $N$ mesoscopic morphogenesis models must be simulated in parallel at each macroscopic time step. On the contrary, macroscopic interactions are efficient to implement as they are based on matrices computations. The model is implemented in \texttt{scala} and integrated into a broader library for spatial sensitivity analysis and dynamical models for settlement systems, namely the \texttt{spatialdata} library \cite{raimbault2020scala}. This library in particular already includes implementations of the models used in \cite{raimbault2020indirect} and \cite{raimbault2018calibration}, but also of other urban interaction models such as \cite{favaro2011gibrat}. The relative large number of parameters and output indicators imposes the use of appropriate model exploration methods to get a grasp of model behavior. Therefore, the parameter space is explored with the OpenMOLE model exploration software \cite{reuillon2013openmole}. Standalone source code for the model is available on an open git repository at \url{https://github.com/JusteRaimbault/UrbanGrowth-model}, while the repository for results is at \url{https://github.com/JusteRaimbault/UrbanGrowth}. Simulation data used in this paper is available on the dataverse repository at \url{https://doi.org/10.7910/DVN/IRHMQK}.

\subsection{Statistical consistency}

% one factor sampling
%(min around 1 for 3 indicators, above 5 for 2 and around 0.5 for 2)}

The first numerical experiment is aimed at checking the statistical consistency of model output across stochastic repetitions. We run a one-factor-at-a-time sampling on all parameters with 5 steps for each within the domain provided in Table~\ref{tab:parameters}, and 500 stochastic repetitions for each parameter.% check these values
Obtained statistical distribution are shown in Fig.~\ref{fig:fig2} for a sample of 10 parameters. Different indicators exhibit various distribution shapes, but are always unimodal and generally rather localized. This means different parameter values can be discriminated when considering summary statistics. To estimate the role of noise, we compute Sharpe ratios (rate between average and standard deviations) for each indicators and parameter values. Median values of these across parameters are always larger than 1.6, confirming a reasonable influence of noise on model outputs. We run experiments with 50 replications for each parameter value in the following.

\begin{figure}[h!]
	\includegraphics[width=\linewidth]{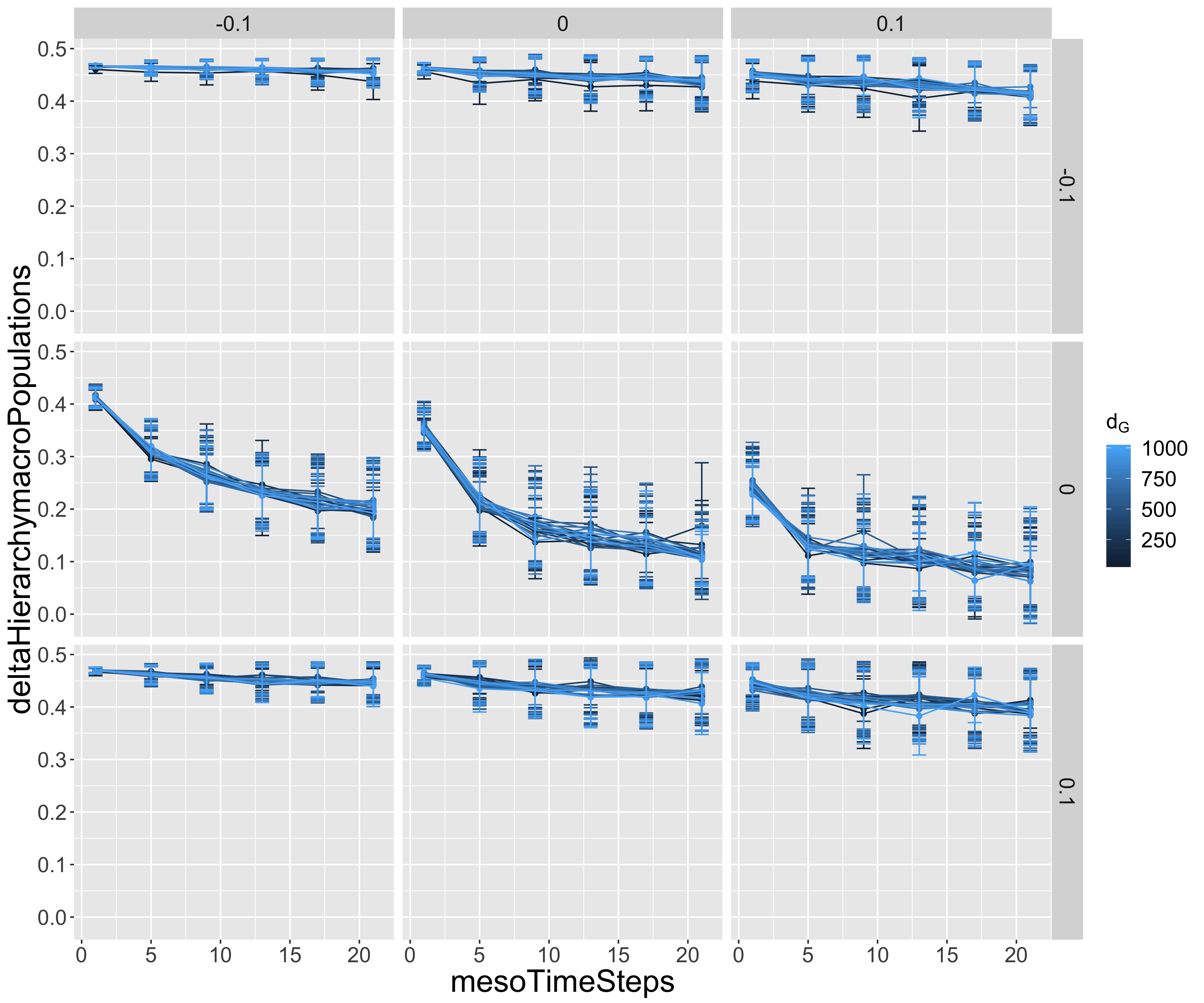}
	\caption{Macroscopic hierarchy of populations as a function of mesoscopic time steps $t_m$, for different values of interaction decay $d_G$ (colour), and of feedback parameters $\delta d$ (columns) and $\delta \beta$ (rows).\label{fig:fig4}}
\end{figure}

\begin{figure}[h!]
	\includegraphics[width=\linewidth]{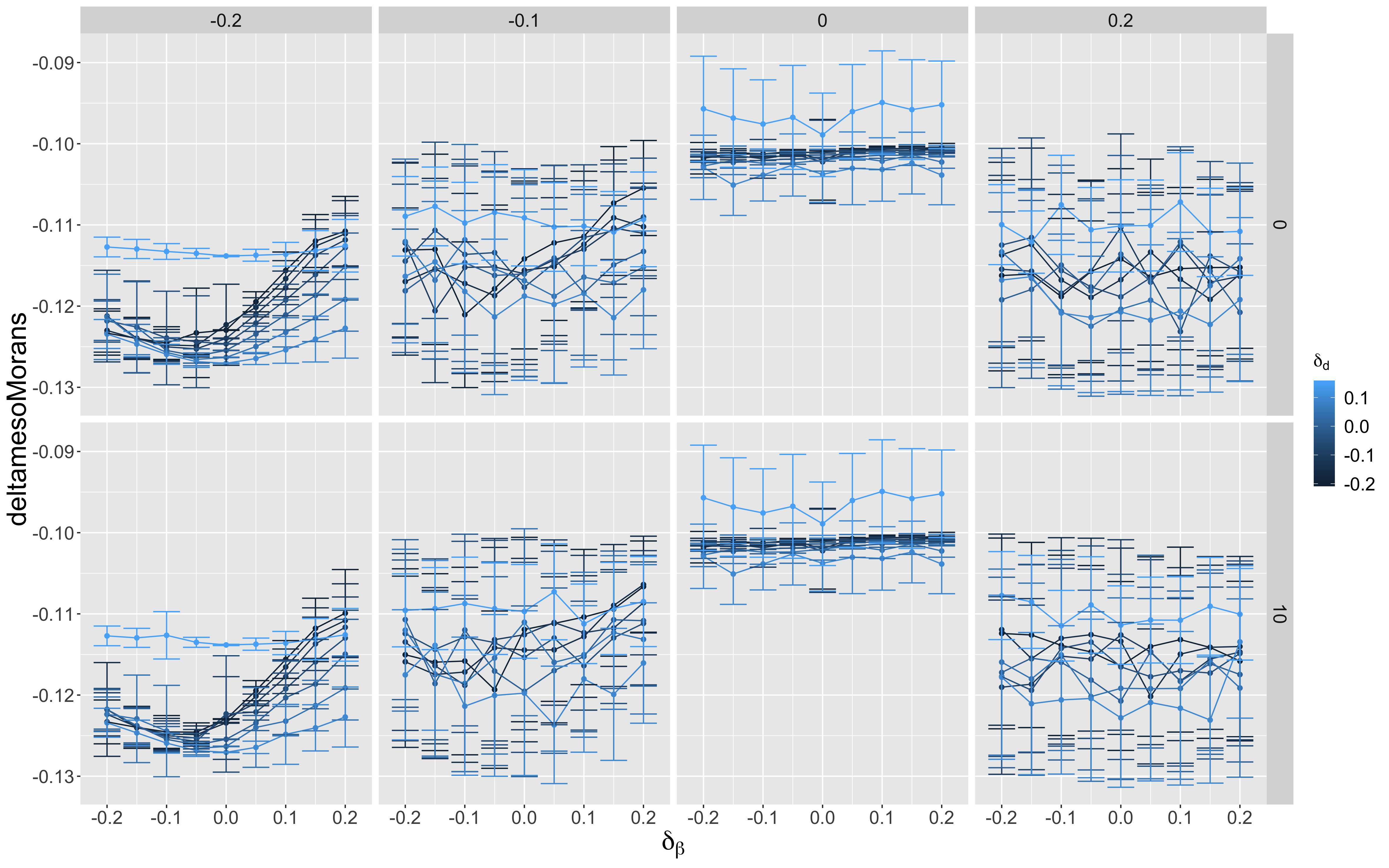}
	\caption{Relative evolution of average mescoscopic Moran, as a function of feedback parameter $\delta \beta$, for different values of feedback parameter $\delta \alpha$ (colour), and of feedback parameters $\delta d$ (columns) and $\lambda$ (rows).\label{fig:fig5}}
\end{figure}

\begin{figure}[h!]
	\includegraphics[width=\textwidth]{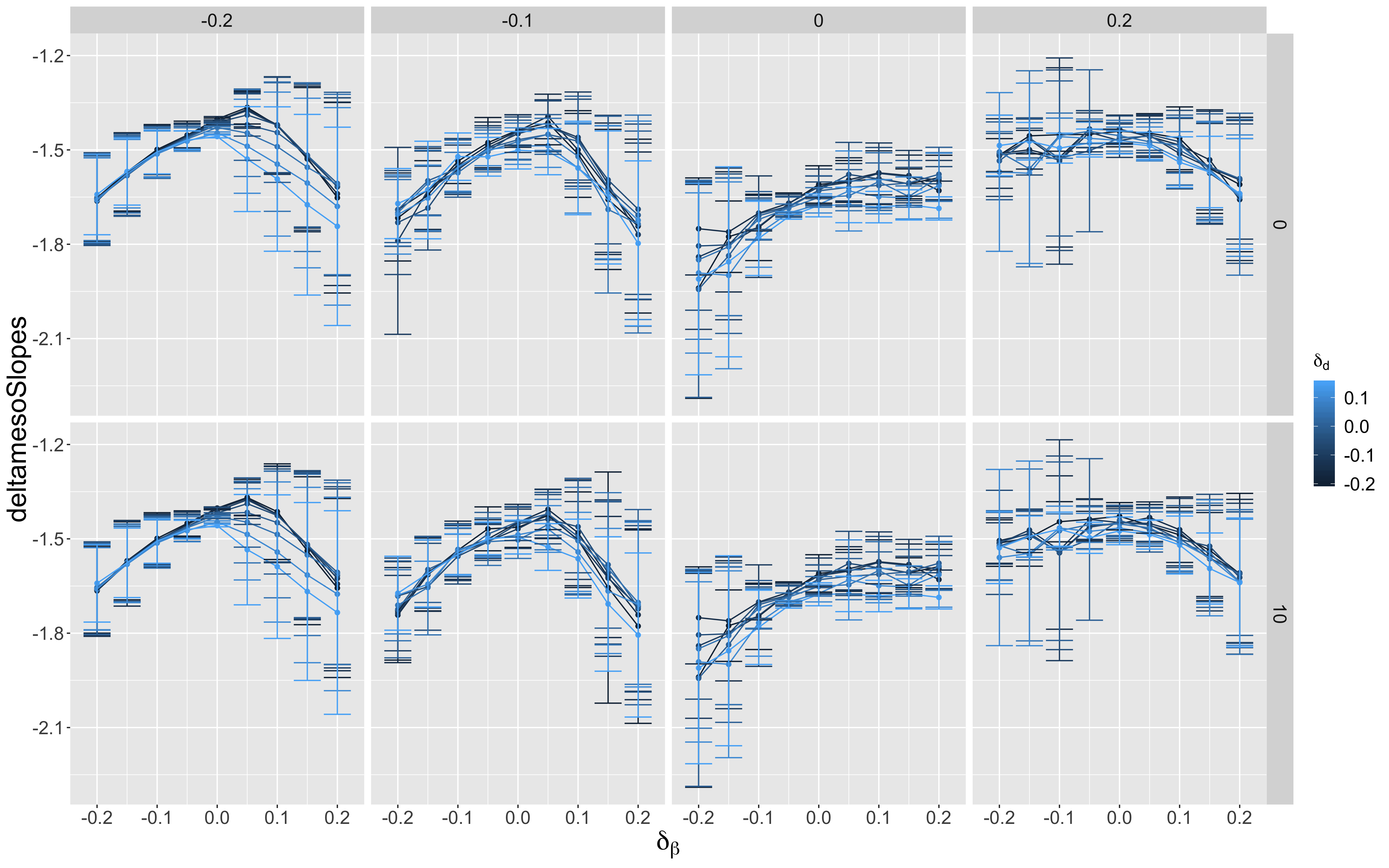}
	\caption{Relative evolution of average mescoscopic population hierarchy, as a function of feedback parameter $\delta \beta$, for different values of feedback parameter $\delta \alpha$ (colour), and of feedback parameters $\delta d$ (columns) and $\lambda$ (rows).\label{fig:fig6}}
\end{figure}

\begin{figure}[h!]
	\begin{center}
	\includegraphics[width=0.6\linewidth]{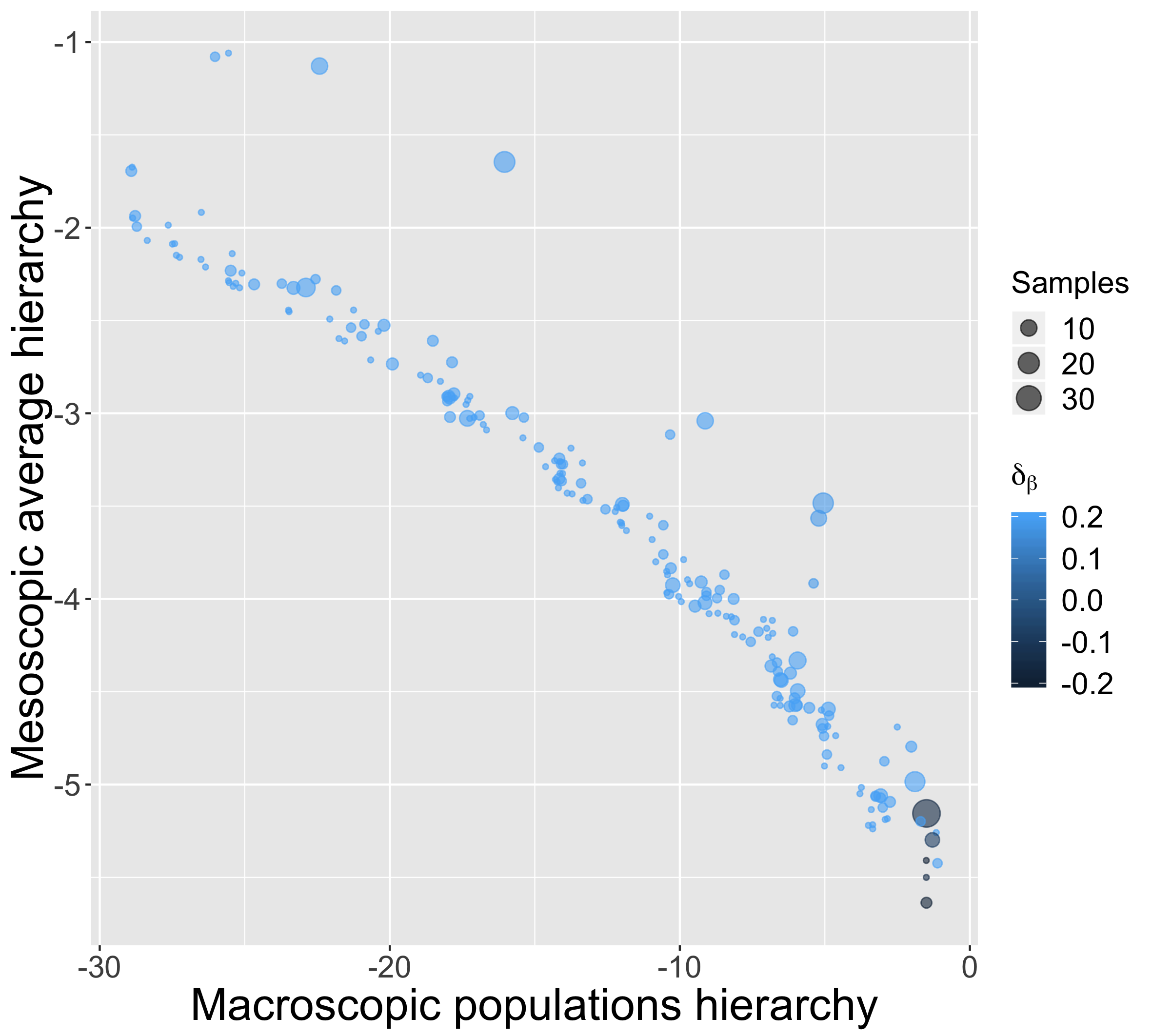}
	\end{center}
	\caption{Pareto front obtained with a NSGA2 genetic algorithm, to minimize simultaneously hierarchy at the macroscopic and mesoscopic scales. Point size gives the number of stochastic samples and colour the value of $\delta \beta$.\label{fig:fig7}}
\end{figure}

\subsection{Grid exploration}

% note : we should control here compare with / without feedback ?
%  -> first thing to check : with/ without the model
%  ~ done when \delta are set to 0 -> could compare systematically to baseline where all feedbacks are deactivated?

% 20190429_193158_MULTISCALE_GRID_GRID.csv
% not the one factor?

% We fix the congestion cost for bottom-up feedback to a rather strong effect with $\lambda = 1$.
% -> in the grid exp?

We then turn to a grid exploration of the parameter space. %precise successive experiments, step values and varied parameters
Results show an important role of the strong coupling between scales. This is shown for example by the variation of macroscopic indicators trajectories when switching from a ``transit-oriented development'' scenario (negative feedback of population growth on diffusion $\delta \beta$) to a ``sprawl'' scenario (positive feedback $\delta \beta$) as shown in Fig.~\ref{fig:fig4} below.

We show in Fig.~\ref{fig:fig3} that both macroscopic and mesoscopic indicators have a U-shaped behavior as a function of the bottom-up feedback parameter $\delta d$. In terms of macroscopic trajectories, we see that hierarchies of accessibilities, closeness centrality and population are maximal in negative value when the mechanism is deactivated, what means that the feedback process moderates the hierarchisation of the global urban system. For the urban form, this feedback also mitigates the aggregation of population.

% 20190506_135221_MULTISCALE_TARGETEDGRID_GRID

% fig 4

The strong coupling of scales, in particular the interaction between bottom-up and top-down feedbacks, has a non-trivial influence on model behavior, as shown in Fig.~\ref{fig:fig4} for the macroscopic hierarchy of populations. This allows investigating the influence of the macroscopic interaction decay and the mesoscopic evolution speed, under stylised scenarios for the top-down feedback. We exhibit interaction effects between $\delta \beta$ and $\delta d$, as quantitative and qualitative behavior as a function of $t_m$ changes across panels. Interaction range $d_G$ has on the contrary no significant effect. We find that increasing $t_m$ decreases macroscopic hierarchy when top-down feedback is deactivated ($\delta \beta=0$), but a cancelling of this effect when it is activated. This may have policy implications in terms of controlling urban sprawl by acting or not on $\delta \beta$, and its influence on the overall urban system.

\subsection{Impact of policy parameters}

We then proceed to a targeted experiment by detailing more the grid for $\delta \beta$ and $\delta \alpha$. The aim is to investigate the influence of policies taken at the level of urban areas, but which can also be constrained by a higher level of governance. We show in Fig.~\ref{fig:fig5} the influence of $\delta \beta$ on urban form. In particular, the average variation of Moran index captures the centralisation of urban areas. This centralisation exhibits a minimal value when $\delta \alpha$ is low, consistent across values of $\lambda$. This implicates that when decentralisation policies are implemented by acting on aggregation ($\delta \alpha$), the policy on urban sprawl will have an unexpected effect on centralisation. This is an illustration of complex and possibly negative interaction between policies.
% influenced by upward feedback? No, rather stable when \lambda = 0 -> why? - check influence of lambda

When investigating population hierarchy the mesoscopic scale, i.e. inequalities of population distribution within the urban area, we find in Fig.~\ref{fig:fig6} a U-shaped behavior as a function of $\delta \beta$ for negative $\delta \alpha$ values, but has a plateau for positive values of $\delta \alpha$, showing also an interaction effect between policies. In terms of hierarchy, a maximal negative value may be wanted to limit inequalities (slopes are negative), and this optimum is reached for a slightly positive value of $\delta \beta$ (left columns in Fig.~\ref{fig:fig6}). This means that sprawl may be needed in some configurations to optimise this objective.

\subsection{Multiscale optimization}

We finally turn to an optimization objective to investigate if compromises have to be made when targets are simultaneously aimed for at different scales. We therefore use a NSGA2 algorithm implemented in OpenMOLE, with a population of 200 individuals, run for 100,000 generations. Objectives are the average mesoscopic hierarchy which can be minimized for more compact and thus sustainable cities, at the macroscopic population hierarchy. Convergence is reached in terms of hypervolume and points in the Pareto front.

We show in Fig.~\ref{fig:fig7} the Pareto front obtained after convergence of the algorithm. This confirms that the two objectives are indeed contradictory and that compromises between scales have to be made when trying to optimise both simultaneously. Most points correspond to policies favouring sprawl, at the exception of a small set of points with low macroscopic hierarchy values, which discourage sprawl ($\delta \beta < 0$). Systems with the most equalities between cities within the system of cities correspond to the most hierarchical urban form, implying inequalities at the mesoscopic scales.

\section{Discussion}

\subsection{Theoretical and practical implications}

Our model effectively captures an interaction between downward and upward feedback. In that sense, it illustrates weak emergence in its proper sense \cite{bedau2002downward}. This suggests this view can be effectively operationalised into models of urban dynamics.

In terms of model complexity and in relation to urban complexity in itself, we show that coupling simple models (each having only three parameters) already yields a complicated and complex simulation model. It also requires additional ontologies to achieve the coupling (in our case the policy processes on one side and the congestion-performance process on the other side). Therefore, we suggest this is additional evidence for a necessity of complexity and simulation models to understand urban complexity. This implies in some way the failure of reductionist epistemologies, at least to grab the multi-scalar nature of urban systems.
% 'suggest in a way failure of reductionist empistemology (in the sense of a la Barthelemy) - at least for grabbing the multi-scalar nature of systems' (seems a tautology though

Although a stylised and highly simplified model, we were already able to integrate parameters linked to policies. Indeed, governance processes are intrinsically multi-scalar \cite{liao2017ouverture}. Therefore, developing such models is a first step towards a progressive integration models for policy and sustainable management of territories.

\subsection{Developments}

%\subsection{Multi-modeling and concurrent processes}

This model is only a first structural sketch with very restrictive assumption, in particular regarding the downward and upward feedbacks on submodel parameters. There may be no link between urban form and global insertion, or it may be due to other processes, be expressed as an other functional form. An important stage before shifting to robust knowledge will consist in (i) reviewing and making a typology of such potential processes across scales; (ii) including most in a multi-modeling fashion to compare possible concurrent mechanisms.

Future work also implies the use of more elaborated model exploration and validation methods. For example, using the diversity search integrated in OpenMOLE \cite{reuillon2013openmole} would allow unveiling the diversity of possible future urban trajectories, and possibly unexpected novel pathways towards sustainability, finding for example regimes with the strongest effect of feedback parameters on indicators related to sustainability.
% spatial sensitivity - influence of synthetic system of cities

Regarding an application of the model on real systems of cities, a calibration on empirical trajectories would be needed. However, such data is difficult to obtain, and some investigations should be made on the relevant calibration targets (in particular at which scale).
% parametrization on real systems; possibly calibration \cite{raimbault2019ilus}

%We also did not include explicitly transportation networks in this model.

Our model is a first step towards multi-scalar models of urban dynamics which effectively include a strong coupling between processes at different scales. While remaining highly stylised, we showed an important role of the coupling and unveiled for example interaction effects between different types of policies. This type of approach will need in the future to be further developed, towards multi-scalar models for sustainable policy making \cite{Rozenblat2018}. This works paves the way for such more complex models.

% conclusion not needed
%\section{Conclusion}
%This paper provides a first step towards strongly multi-scalar models to capture urban complexity towards policy models. 
%Towards integrative models and theories for urban systems \cite{raimbault2019methods}

\section*{Acknowledgments}

The author acknowledges the funding of their institutions and the EPSRC project number EP/M023583/1. Results obtained in this paper were computed on the vo.complex-system.eu virtual organisation of the European Grid Infrastructure (http://www.egi.eu). We thank the European Grid Infrastructure and its supporting National Grid Initiatives (France-Grilles in particular) for providing the technical support and infrastructure.

\end{document}